# Temperature-dependent permittivity of annealed and unannealed gold films

Po-Ting Shen[1], Yonatan Sivan[2], Cheng-Wei Lin[3], Hsiang-Lin Liu[3], Chih-Wei Chang[4], and Shi-Wei Chu[1,5,*]

[1]*Department of Physics, National Taiwan University, 1, Sec. 4, Roosevelt Road, Taipei 10617, Taiwan (R.O.C.)*
[2]*Unit of Electrooptics Engineering, Faculty of Engineering Sciences, Ben-Gurion University of the Negev, P.O. Box 653, Be'er Sheva, Israel 8410501*
[3]*Department of Physics, National Taiwan Normal University, 162, Sec. 1, Heping E. Rd, Taipei, Taiwan (R.O.C.)*
[4]*Center for Condensed Matter Sciences, National Taiwan University, Taipei 10617, Taiwan (R.O.C.)*
[5]*NTU Molecular Imaging Center, 81, Changxing St., Taipei 10672, Taiwan (R.O.C.)*
[*]*swchu@phys.ntu.edu.tw*

**Abstract:** Due to local field enhancement and subwavelength confinements, nano-plasmonics provide numerous novel applications. Simultaneously, as an efficient nanoscale heat generator from inherent absorption, thermo-plasmonics is emerging as an important branch. However, although significant temperature increase is involved in applications, detailed characterization of metal permittivity at different temperatures and corresponding thermo-derivative are lacking. In this work, we extract the permittivity of gold from 300K to the annealing temperature of 570K. By comparing annealed and unannealed films, more than one-order difference in thermo-derivative of permittivity is revealed, resulting in unexpectedly large variation of plasmonic properties. Our result is valuable not only for characterizing extensively used unannealed nanoparticles, but also for designing future thermo-nano-plasmonic systems.

## 1. Introduction

Nano-plasmonic systems have been intensively studied in recent decades due to their unique potential for local field enhancement and subwavelength confinement, that are promising for a wide variety of applications [1, 2]. In the early stage of development, the inherent absorption in the metal was thought to be a substantial obstacle towards real-life applications. However, in recent years the applied plasmonic researches exploit the absorption as means to generate heat on the nanoscale [3]. The temperature scale of this new sub-field, thermo-plasmonics, starts from photothermal imaging [4], through cancer treatment [5], super-resolution microscopy [4, 6, 7], plasmonic photovoltaics [8, 9], water boiling, and super-heating [1, 10, 11], up to thermo-photovoltaics [8], solar thermo-electric generators [12], plasmon-mediated photocatalysis [9], plasmon-assisted chemical vapor deposition [13] and heat-assisted magnetic recording [14], which may involve temperatures higher than 2000K [15].

Such studies obviously require knowledge of temperature-dependent metal permittivity. However, it turns out that experimental measurements of such dependencies are rather scarce and there is no complete and correlated set of data from room temperature up to the melting temperature. For example, the old data of [16], extracted from spectroscopic data of a heated gold film (295 - 770K), specifies the absorbance (from which the imaginary part of the permittivity can be extracted), but not the real part of the permittivity. Compared to the well-recognized data of [17], the absorbance value in the former study is rather high, possibly due to high surface roughness in their samples.

In a recent study [18], the temperature dependence of absorption from gold nanospheres in silica was measured. Although detailed extinction spectra is shown over the widest temperature range studied so far (up to ~1200K), their results cannot unambiguously extract the real and imaginary parts of the gold permittivity, critical to study other geometries.

To date, the most systematic study of the temperature dependence of gold permittivity was performed in two recent works [19, 20]. However, instead of showing the values of the permittivity itself, only values of the thermo-derivative were provided in [19]. Moreover, a rather rough estimate of the thermoderivative was provided, based on only two temperatures (300 and 430K). On the other hand, better temperature resolution is provided in [20], but only for a narrow spectral range (700-900nm) and temperatures beyond the boiling point. No *in-situ* analysis of annealing effects was considered.

It is well known that increasing temperature induces changes in the morphology of gold due to annealing. Surprisingly, no morphological information was included in previous temperature dependent permittivity studies. Although an irregularly small thermo-derivative is observed at 460K in [20], no discussion about annealing was given. It is known that annealing significantly affects plasmonic effects [21, 22]. However, since most plasmonic nanostructures are fabricated without annealing, it is critical to examine the variation of fundamental physical property of gold before and after annealing.

In this work, *in-situ* temperature dependent permittivity of gold before and after annealing was studied from room temperature to annealing temperature, with 10K steps. We found more than one order difference in the thermo-derivatives, for the first time to our knowledge. In addition, we found an interesting anomaly that the imaginary permittivity of gold film increases after annealing, possibly due to the scattering effect from the gold grain. Our work will serve as a valuable reference to characterize gold nanophotonic devices.

## 2. Experimental setup

The gold film was made by electron beam evaporation of gold with 99.99% purity onto a silicon nitride substrate with 5nm titanium as an adhesion layer. Its thickness is 130±16 nm, which should be enough to be treated as bulk [23]. For unannealed sample, the deposited film is directly used. For annealed sample, the gold film is placed at 600K for 15 minutes, under $10^{-3}$ atm, and then cooled off to 300K. The temperature dependent permittivity is acquired with a spectroscopic ellipsometer (M2000 U, J.A. Woollam Co, NE) from 190 to 1600 nm.

## 3. Results

### 3.1 Permittivity at room temperature

Fig. 1(a) and 1(b) shows the real part and imaginary part of gold permittivity measured at 300K. The data from annealed and unannealed films are plotted as red and blue lines. The morphology of gold film before and after annealing is shown in Fig. 1(c) and (d), respectively. Obviously, the grain size greatly increases in Fig. 1(d), manifesting the effect of annealing.

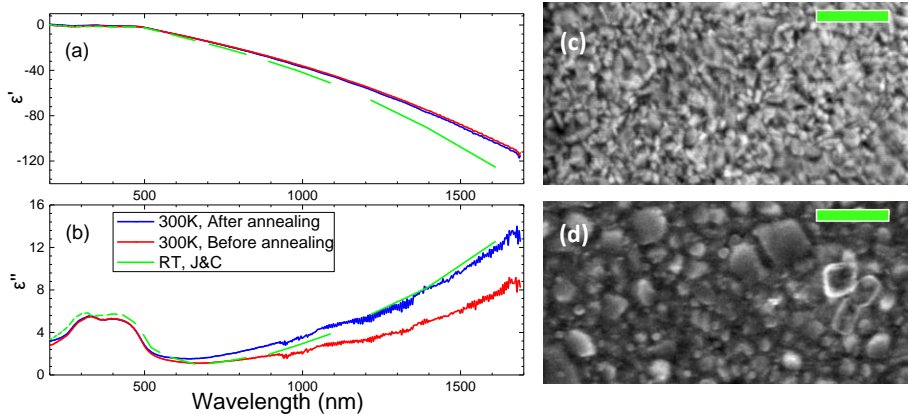

Fig. 1. (a) The real part ε' and (b) imaginary part ε" of permittivity of gold at room temperature. Morphology (c) before annealing (d) after annealing. Scale bar: 400nm

Our permittivity results are in good agreement with the generally recognized work of [17], shown as green dashed lines. Interestingly, in the visible range, the permittivity (both real and imaginary) is not affected significantly by annealing. In the near-infrared region above 1000 nm, slight deviations, in both real and imaginary parts, are found between our data and [17] before annealing. For the real part, annealing does not affect the correspondence much, but for the imaginary part, after annealing, it becomes well-matched to [17] again.

*3.2 Temperature-dependent permittivity of unannealed gold*

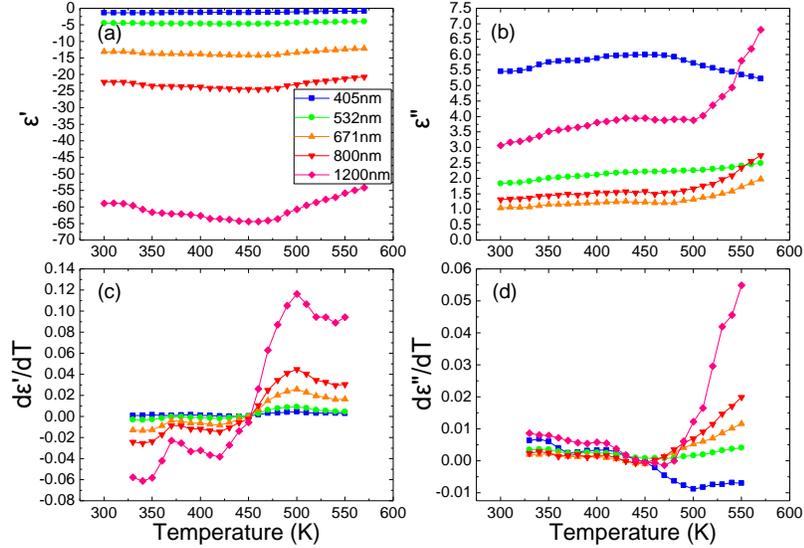

Fig. 2. (a) The real part ε' and (b) imaginary part ε" of permittivity with unannealed gold films. (c) and (d) are the thermo-derivatives of ε' and ε".

By increasing temperature gradually on the unannealed film, remarkable variation of gold permittivity throughout the spectral domain is observed, as shown in Fig. 2. For the real part in Fig. 2(a), the data below 520nm exhibits only small variation. The variations are more and more significant for increasing wavelength, but all exhibit the same trend of decrease and then increase above 460K, reflecting the onset of annealing effect. On the other hand, for the imaginary part in Fig. 2(b), monotonical increase is observed for wavelengths above 520 nm; while for 405nm, the trend is increase and then decrease above 460K. The largest variation is found at 1200nm, whose imaginary permittivty increases by 133% from 300 to 570K.

To highlight the trend of variation, the real and imaginary parts of thermo-derivatives are given in Fig. 2(c) and (d), respectively. The real parts are negative at all wavelengths at low temperatures; however they become positive above 460K, with a peak at around 500K. For the imaginary parts, all decrease from positive at low temperature, and revert at 450 – 500K. Here, only wavelength below 520nm exhibits negative derivative at high temperatures. By dividing the thermo-derivatives to the permittivities, the relative change of imaginary part is much larger than real part.

*3.3 Temperature-dependent permittivity of annealed gold*

Figures 3(a) and (b) show the permittivity of the annealed gold film with increasing temperature. Compared to Fig. 2, the variation is much smaller and smoother. For both the real and imaginary parts in the visible band, linear trends are observed below 450K, justifying the result in [19]. In Fig. 3(a), for 405nm, the overall permittivity variation is 12%; while for the rest wavelengths above 520nm, the permittivity variations are less than 5%. On the other hand, in Fig. 3(b), for 405nm, the permittivity variation is less than 6%; for other wavelengths, the variations are in the range of 20 - 40%, much smaller than the unannealed film.

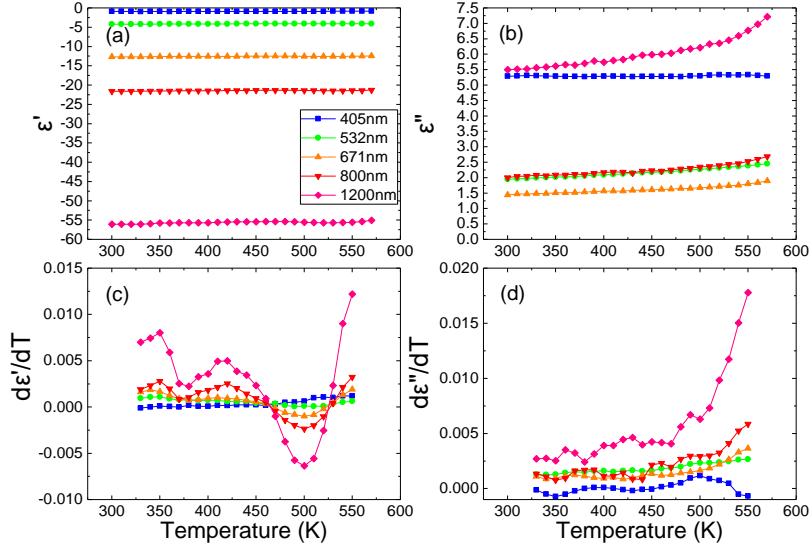

Fig. 3. (a) The real part ε' and (b) imaginary part ε" of permittivity with annealed gold films. (c) and (d) are the corresponding thermo-derivatives of ε' and ε".

The corresponding thermo-derivatives of real and imaginary parts are drawn in Fig. 3(c) and 3(d), respectively. Except 405 nm, all wavelengths exhibit similar trends. The real parts decrease at low temperature, and start to increase above 500K. The imaginary parts are positive and increase monotonically. Comparing to Fig. 2(c) and 2(d), the thermo-derivatives of the annealed film are nearly one order smaller than the unannealed film.

## 4. Discussion

In this work, we find from the unannealed film that not only the room-temperature permittivty is different from the annealed film, but the thermo-derivative is also much larger. Since most metallic nanoparticles are fabricated in chemical solutions [24], they are in nature multi-crystalline, i.e. unannealed. It is apparently critical to use correct permittivity data for plasmonic calculations.

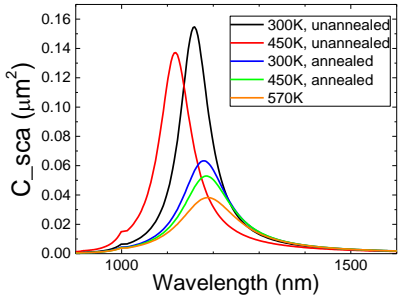

Fig. 4. Temperature and annealing dependencies of scattering from a gold nanoshell (diamond core radius 50nm, gold shell thickness 5nm).

To highlight the significance, Fig. 4 shows the calculated scattering from a gold nanoshell, resonant near 1200 nm in the biological transparent window [25]. More than 150% difference is found between the annealed and unannealed spectra at 300K. It is even more dramatically to consider the temperature dependent resonance. For the unannealed particle, more than 400% variation is seen between 300K and 570K, while the annealed one exhibits less than two-fold variation. To the best of knowledge, although significant temperature change is assumed in previous calculations for thermo-plasmonic applications [1, 2], such permittivity variation has never been considered. Based on our finding, the factor of temperature-dependent permittivity and annealing need to be included in future thermo-plasmonic studies.

It is interesting to note an unusual phenomenon in Fig. 1(b): higher imaginary permittivity after annealing, against the prediction of Drude-Sommerfeld model. Two recent works [26, 27] that compared morphological information versus permittivity also find larger imaginary permittivity from a gold film with large grains. We speculate that the increased loss is due to the scattering effect on gold grains. It is well known that as the diameter of scatterer is much smaller than wavelength, Rayleigh scattering dominates, and the scattering intensity is roughly equal in all directions. On the other hand, when the diameter increases, Mie scattering takes over, and the scattering is more concentrated in the forward direction. As shown in Fig. 1(c), before annealing, the grain size on the film is much smaller than incident wavelength. In Fig. 1(d), after annealing, the grain size approaches wavelength. In the ellipsometry setup, reflection from the gold film is measured to determine the imaginary permittivity. With large grains, more light scatter in the forward direction, and thus less light in the reflection path, resulting higher loss, as well as imaginary permittivity.

Having said this, it is surprising that our imaginary permittivity value of the annealed film fits reasonably well with [17], in which the films were annealed at 423K for 9 hours. The low-temperature and long annealing should in principle create a uniformly annealed film. However, since no morphological information was provided in [17], and our permittivity result fits well to theirs, we suspect that their result in fact represents the permittivity of large grains of gold, i.e. partially annealed gold. To further justify this point, two recent reports [28, 29] with high-quality films showed smaller imaginary permittivity than results from [17].

A final remark is on the adhesion layer. During our sample deposition, a titanium layer facilitated the adhesion of gold, and thus improved the overall film flatness. Nevertheless, when the film was heated to annealing temperature, the adhesion force from the titanium layer might prevent gold from forming a uniform film, resulting in large grains of gold, and thus more scattering in the forward direction. In terms of the thermo-derivative, titanium should not play a significant role during the measurement. Compared to [20], where deposited gold film with a Cr adhesion layer is used, the resulting thermo-derivative of gold is very similar to our result, manifesting the insignificant role of the adhesion layer.

## 5. Conclusion

Thermo-plasmonics is emerging as an important sub-field in plasmonic applications. It is well known that nano-plasmonic properties are dominated by metal permittivity, which is highly dependent on temperature and crystalline structure. Here we provide detailed characterization of thermo-derivatives of annealed and unannealed gold for the first time, and found substantial differences. Conventionally, permittivity of annealed gold is adopted to calculate plasmonic resonance. However, most plasmonic nanostructures are fabricated with unannealed metal, thus by using unannealed permittivity, the plasmonic resonance strength can be very different. Our work not only features the first complete temperature dependent permittivity measurement of annealed versus unannealed gold, but also points out the importance of adopting correct set of permittivity for future thermo-nano-plasmonic studies.

This work was supported by the National Science Council of R.O.C. under Contract No. NSC-102-2112-M-002-018-MY3 and NSC-101-2923-M-002-001-MY3. SWC acknowledge the generous support from the Foundation for the Advancement of Outstanding Scholarship.